\documentclass[twocolumn,showpacs,amsmath,amssymb,floatfix]{revtex4}

\usepackage{graphicx}
\usepackage{dcolumn}
\usepackage{bm}

\usepackage{dcolumn}

\newcommand{\euler}[1]{{\usefont{U}{eur}{m}{n}#1}}
\newcommand{\umu}{\mbox{\euler{\char22}}}

\newcommand{\umum}{\mbox{\euler{\char22}}\textrm{m}{}}

\usepackage{wrapfig}

\usepackage{color}

\usepackage{ulem}

\newcommand{\mymark}[1]{#1}

\begin{document}

\title{Guiding of dynamically modulated signals in \mymark{arrays of
  photorefractive spatial solitons}}


\author{Denis Tr\"ager}
\email{denis.traeger@uni-muenster.de}
\author{Nina Sagemerten}
\author{Cornelia Denz}
\affiliation{Institut f\"ur Angewandte Physik, Westf\"alische Wilhelms-Universit\"at M\"unster, 48149 M\"unster, Germany}







\begin{abstract}
As the formation of spatial optical solitons in photorefractive media
is governed \mymark{by modification} of the refractive index every single
solitons in complex configurations of solitons can act as a single
waveguide for other light beams. 
In this article we demonstrate guiding of amplitude modulated beams in
complex configurations of photorefractive solitons carrying
information and we analyze the received signal in the kHz frequency
range. 
The possibility of data transmission combined with waveguide 
couplers opens the route to all-optical networks.
\end{abstract}


\maketitle

\section{Introduction}

The fascinating properties of photorefractive spatial
solitons lead to many possible applications in optical
telecommunications. 
The formation of these solitons is based on the modification of the 
material's refractive index by incident light.
As a direct consequence this local refractive index change can be
exploited especially to guide beams of other wavelengths. 
In combination with interaction scenarios in large complex
configurations of solitons the waveguiding features are essential for 
the usage in telecommunication applications. 

The basis for analyzing the suitability of optical spatial 
solitons in photorefractive media is the study of their fundamental
properties (for an overview see \cite{WAPT+2002,segevopn}). 
Here, especially mutual interaction of single solitons like Gaussian
and also higher order modes has been studied thoroughly
\cite{WAPT+2002}.
Because all-optical interconnection devices are of special interest, 
the waveguiding features play a major role so that single
waveguides~\cite{SSS1996} and waveguide couplers had first to be
demonstrated~\cite{PDSK2002}. 
Before, waveguiding features of single Gaussian mode solitons have
been investigated~\cite{SCMSx1997} and higher order solitons like
vortices have been predicted~\cite{CMALx2000}.
\mymark{
Also, more complex structures like arrays of photorefractive
solitons~\cite{PTSBx2003,TSSDx2003} and other materials than
photorefractive crystals like organic photorefractive glasses or
nematic liquid crystals~\cite{ASCOM2005,PDADx2000} have been shown to
guide light of different wavelengths. 
Lattices of solitons and more complex photonic structures are even
able to exhibit band structures and can be investigated by excitation
of Bloch waves (see \cite{CLS2003,TFNSx2006} and references therein).  
}

Guiding of visible and infrared laser beams using photorefractive
solitons as waveguiding channels demonstrate their principle
suitability for all-optical data transmission and information
processing. 

Exploiting the parallelism of waveguide arrays, many waveguide channels
can be addressed in parallel and transmit information. 
However, the transmission of information in waveguides is always
closely connected to the adequate temporal modulation of the guided
beam. 
Therefore, in order to use spatial optical solitons for information
transfer, the transmission characteristics and  loss behaviour of a
spatial soliton waveguide needs to be known. 
Consequently in this article we demonstrate successful transmission of 
amplitude modulated signals through waveguiding photorefractive
solitons in a large \mymark{array}. 
In this contribution, we demonstrate to our knowledge for the
first time the successful transmission of high-speed amplitude modulated
real data signals through waveguides that are built by photorefractive
solitons in large soliton arrays.
To achieve this we have implemented a setup that is able to modulate, 
transmit and demodulate a guided laser beam using an amplitude
modulation technique.

\section{\label{sec:exp}Experimental realization}
The experimental realization of a complex configuration of
photorefractive solitons used for information transport is based on
earlier experiments~\cite{PTSBx2003,TSSDx2003}. 
\begin{figure}[b]
  \includegraphics[width=\linewidth]{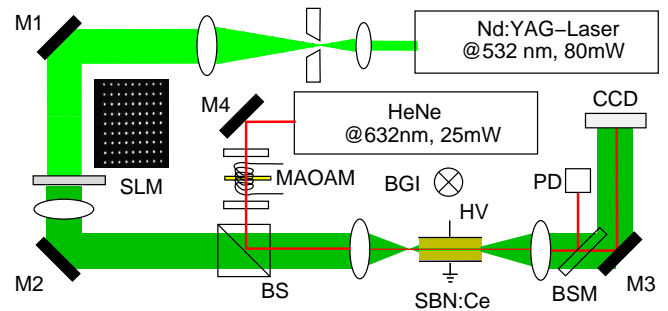}
  \caption{\label{fig:epsart}Experimental setup for preparing 
    \mymark{array} of optical spatial solitons as waveguides and the
    transmission of modulated signals
    (SLM: spatial light modulator/amplitude mask, M: mirror, 
    BS: beamsplitter, BSM: beam splitting mirror, MAOAM:
    magneto-optical amplitude modulator, HV: high voltage, 
    BGI: backgrund illumination, CCD: CCD-camera, PD: photodiode)} 
\end{figure}
The setup we are using for the experiments described in this article
is shown in figure~\ref{fig:epsart}. 
The output beam of a 532\,nm frequency-doubled Nd:YAG-laser with an output
power of 80\,mW is spatially filtered and broadened to a diameter of
about 30\,mm. 

We are imprinting the \mymark{array} configuration onto this beam with
an amplitude modulator - a liquid-crystal diplay (LCD, 640x480 pixels)
in amplitude mode driven by a computer graphics card. 
One example of a rectangular \mymark{array} configuration with 9x9
single beams is shown in the inset of figure~\ref{fig:epsart}. 
Moreover, the system allows to imprint arbitrary images on the beam,
and we have therefore a very flexible tool for generating more or less
complex spatial soliton configurations.

The spatially modulated beam is then demagnified and imaged onto the
front face of the photorefractive Cerium-doped
Strontium-Barium-Niobate crystal (SBN:Ce) we are using.  
To address the strong photorefractive nonlinearity the light is
extraordinarily polarized in order to use the high photorefractive
coefficient along the material's c-axis.
A white light background illumination adjusts the dark conductivity of
the material to prepare the conditions for soliton generation. 
With the help of an external electric bias field 
applied along the c-axis each of the \mymark{array} beams is able to
produce a single solitonic waveguide. 
The distance of the single beams is chosen suffiently large to avoid
interaction between solitons.

Experiments on guiding beams of other wavelength in photorefractive
solitons have been shown in~\cite{PDSK2002,PTSBx2003,TSSDx2003}. 
In different soliton \mymark{array} configurations guiding of
unmodulated beams with wavelengths 633\,nm and 1.55\,\umum{} has already
been studied.  
Using this as starting point we have extended our setup to modulate
the guided light beams to simulate data transmission in such a system. 

The amplitude modulator we are using is based on a magnetooptic crystal
(Yttrium-iron-garnet (YIG) film with Bismut\footnote{Kind donation of
  Prof. Dr. Horst D\"otsch at Universit\"at Osnabr\"uck, Germany.}) which 
responds with Faraday rotation depending on an external magnetic
field. 
Typical modulation frequencies of this material span up to Gigahertz
range.
\begin{wrapfigure}[16]{l}{5cm} 
  \includegraphics[width=\linewidth]{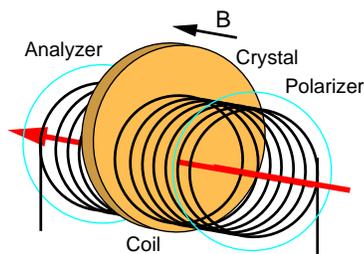}
  \caption{\label{fig:maoam}
    Schematic of the magneto-optical amplitude modulator setup. 
    The red arrow indicates the beam propagation direction.
  }
\end{wrapfigure}
Modulation of the external magnetic field is performed using a long
copper coil with about 20 turns that induces a relatively
homogeneous field on its inside. 
The crystal is positioned in the center of the coil to ensure
that the magnetic field is perpendicular to the crystal's surface%
\footnote{Depending on this field the polarization is turned with
  typical values for the Faraday rotation constant of about
  1000$^\circ$cm$^{-1}$ at 1.3\,\umum{} and several
  10000$^\circ$cm$^{-1}$ for visible wavelengths. 
  The thin crystals created by liquid phase epitaxy have typical
  thicknesses of several 10\,\umum{}  which allows values for the 
  resulting Faraday rotation near 90$^\circ$ only for relatively strong
  and homogeneous magnetic fields. 
}.
The coil is placed between polarizing and analyzing polarizers to
translate the polarization rotation into amplitude modulation of a
passing light beam (see figure~\ref{fig:maoam}). 
As driver for the coil we use a music-amplifier with a maximum output
power of 2\,W that is connected to a sound card output to permit for 
transmission of computer generated signals\footnote{This
  configuration allows a modulation depth of the 
  transmitted beam of around 30\,\%.}.
A red laser beam derived from a HeNe-Laser at
$\lambda=632\,\mathrm{nm}$ with 25\,mW output power is amplitude
modulated as described and focussed onto one of the beam positions on
the front face of the photorefractive crystal to achieve coupling into
the corresponding solitary waveguide. 

The guided and therefore transmitted signal is detected using a
photodiode that receives 90\,\% of the light due to the semitransparent
mirror in front of the camera where the other 10\,\% are used for
the imaging. 
We measure only the AC part of the current, the DC part is filtered
with an electrical transducer behind the electrically amplified
photodiode. 
A second amplifier matches impedance and amplitude of the electrical
signal to the input of the computer's analog-to-digital-converter.

\section{\label{sec:res}Experimental results}

\subsection{General characterization}

\begin{figure}[htbp]
  \includegraphics[width=\linewidth]{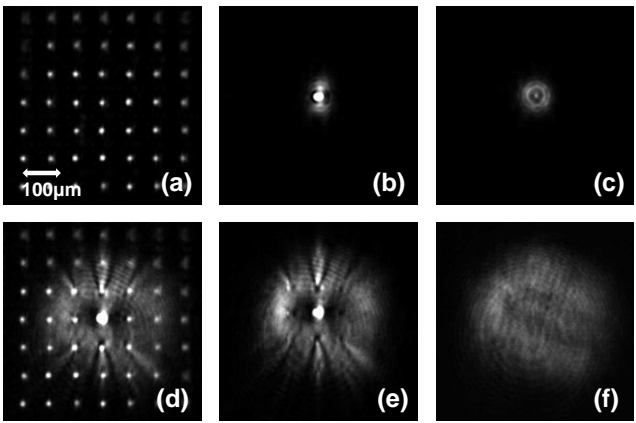}
  \caption{\label{fig:signal}%
    \mymark{ 
    Spatial filtering at crystals output
    face with a Gaussian beam guided in a 7x7 array. 
    }
    (a) \mymark{Array} and beam together at output face of the crystal.
    (d) \mymark{Array} and beam together at output face of the crystal.
    Guided beam with (b) and without (e) spatial filtering with
    pinhole.
    Diffracting beam after waveguide removal with (c) and without (f)
    spatial filtering with pinhole.
  }
\end{figure}

The first step to successfully transmit data with our setup is to
generate photorefractive solitons in a suitable configuration for
these experiments. 
\mymark{%
A rectangular \mymark{array} of 7x7 solitons is chosen for easy
addressing of neighbouring channels simply by moving the signal beam
in horizontal or vertical direction.  
}%
The \mymark{array} configuration is depicted in
figure~\ref{fig:signal}~(a). 
Because of the typical distance of the solitons of 80\,\umum{} strong
interaction between them is avoided, so that every beam generates an
independent waveguiding channel created with an average light power of
about 110\,nW. 
\mymark{%
The limiting factor of cross-coupling between parallel waveguiding 
channels does not depend on the temporal dynamics of the signal but
on neighbouring waveguide distance. 
Therefore, in this article we investigate parallel beams with
sufficiently large distance to avoid cross-coupling between channels
in accordance to \cite{PSTD2002,PTSBx2003}.
}%
The typical diameter of a single soliton is about 25\,\umum.
Nonlinearity is adjusted by an external electric field of 1\,kV/cm
applied along the c-axis of the crystal. 
Background illumination, beam power and external electric field are
adjusted in order to allow for stable photorefractive soliton
generation. 

\mymark{
In agreement with earlier experiments, when large arrays of solitons
are produced, the Gaussian shape of the beam illuminating the LCD 
and inhomogeneities of both the modulator and the crystal have 
to be considered. 
While the last two have a negligible influence in our case, the
Gaussian light intensity distribution leads to larger time constants
to reach stationary state for the solitons at the borders of the array
compared with central ones. 
But due to saturation effect the same refractive index modulation is
induced and therefore the same waveguiding properties are resulting. 
In order to confirm this, we first tested beam and signal transmission
in  different waveguiding channels successively and observed similar 
behaviour. 
}

To make sure we only receive and analyze the light that is being
guided in the chosen photorefractive soliton we use an iris diaphragm 
($\simeq 30\,\umum$ diameter) placed directly at the output face of
the crystal. 
This simulates a small aperture comparable to a large fiber input or a
taper used for further transmission of the beam. 
In figure~\ref{fig:signal} we have displayed pictures of the output
face with (b-c) and without (d-f) this spatial filter.

\mymark{%
The time-resolved experiments themselves have been conducted using
different pieces of music as input signal for amplitude modulation. 
}%
The characteristics of this are typical for todays digital music
systems. 
We have a frequency range from 20\,Hz to 22\,kHz (for the one stereo
channel that we use) together with a dynamic range of 16 bit per
sample.  
This gives a theoretical pulse length (when considering digital
data transmission) of about 45\,\umu{}s and a data rate of 22 kbit/s.
With our setup we cannot use the whole dynamic range of the original
signal but can easily discriminate and analyze the different features
of the system.

\subsection{Signal transmission analysis}

To investigate the transmitted signal through one of the 
waveguides in the \mymark{array} we start recording the modulated
light 
detected by the photodiode when the signal transmission begins. 
It is digitized by means of the analog-digital-converter in our
computer. 
\begin{figure}[htbp]
  \includegraphics[width=0.93\linewidth]{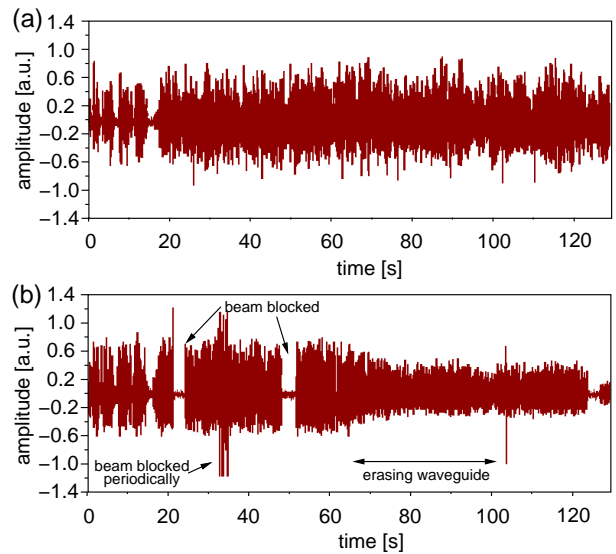}
  \caption{\label{fig:music}%
    Transmission of dynamic amplitude modulated signal. 
    (a) shows the original piece of music for comparison, 
    (b) the transmitted and then recorded version for analysis:
    The beam is blocked at 26\,s to determine the signal-to-noise 
    ratio. 
    The dynamic range is studied at 37\,s with periodic blocking of the 
    beam.
    Erasing of the waveguide starts at 61\,s and results in a reduced 
    recorded amplitude (by about 50\,\%).
  }
\end{figure}
Figure~\ref{fig:music}~(a) shows the amplitude plot of the original
piece of music as reference. 
In comparison the transmitted and recorded signal that undergoes
several tests to determine the signal-to-noise-ratio and some other
properties is depicted in figure~\ref{fig:music}~(b).

In order to test the quality of the transmission in the spatial
soliton, we performed several tests during transmission. 
The signal quality is determined by two main factors: the
modulation depth that needs to be maintained as present in the original
signal and the transmission of all frequencies present in the signal. 
Figure~\ref{fig:music} shows some of the basic investigations related
to that question.  
Properties like noise and modulation transfer of the system are
considered.  

At the beginning the strong even visible amplitude modulation during
the first part (0\,s to 20\,s) is directly related to the artistic
work.  
The noise level present in the system without signal transmission is
primarily determined by the quality of the transmission setup which
incorporates some internal and external noise sources. 
The photorefractive crystal and the solitary waveguides themselves are
not creating any measurable additional noise. 
External light sources and power supplies operating in the same
frequency range show a much stronger influence on the signal quality. 
The basic noise level without data transmission is depicted in
figure~\ref{fig:music}(b) for example at the end starting from 124\,s
to 127\,s. 

From 22\,s to 25\,s the beam was simply blocked and unblocked again. 
This has been done to be able to differenciate between noise
originating from the complete system including modulator and red laser
in the first case from the detection unit in the second case. 
We found the difference to be negligible. 
The signal-to-noise ratio of this system is calculated to be 12\,dB
which is completely sufficient to study the fundamental features of
the investigated system.

A second test here shows several spikes in the amplitude due to
repeatedly blocking of the laser beam and demonstrate signal flanks
with maximum amplitude (``cracking noise'', see figure~\ref{fig:music}
from 37\,s to 38\,s).
This gives us information on the dynamic range of the system that is
calculated to be about 14.5\,dB above the intrinsic noise.

In the right half of figure~\ref{fig:music}~(b) the amplitude plot of
the second part of the experiment is depicted. 
The transmission of the signal has been conducted through an already 
prepared solitary waveguide as a part of an array of solitons as
before (see figure~\ref{fig:signal}~(a)).
To test the waveguiding features and the stability of the waveguide we
start erasing the refractive index modulation by applying a strong
white light illumination to the crystal (starting at 60\,s). 
This leads to vanishing of the waveguiding properties and therefore to
a broad diffracted Gaussian beam at the crystal output face. 
Because of the spatial filtering with the small aperture at the
crystal output face (shown in figure~\ref{fig:signal}) the 
amplitude of the received signal at the detector is reduced to about 
50\,\% during the erasure process (completed at 85\,s).
Then less light is passing the filtering hole due the to broadened
beam.

In agreement with earlier results~\cite{PD2001} the coupling
efficiency for the 632\,nm beam in a solitary waveguide amounts 
to about 25\,\% without any special optimization like
beam-to-numerical aperture matching. 
For the probe beam we have used a lens with 60\,mm focus length to
achieve a sufficiently small spot on the input face of the crystal
while the working distance is large enough to integrate it into
the existing system. 
\mymark{%
The system allows for parallel transmission of multiple
beams - in principle only limited by the number of waveguides
written. 
Therefore, the single channel data rate can be multiplied with the
number of independent channels. 
Additionally, possible wavelength multiplexing of signals in the same  
waveguide has not been investigated experimentally in this work.
}%

\mymark{%
The channel cross-coupling for modulated signals is already determined
by the amount of light exchanged between neighbouring channels in the
unmodulated case. 
Therefore, to minimize cross-coupling this has to be controlled. 
As already discussed in earlier works \cite{PD2001} interaction
strength of solitons depends primarily on distance. 
Coherence and relative phase govern the actual type of interaction
based on the overlapping electric fields of the involved beams. 
In the studied case of coherent in-phase beams propagating in parallel
to form the arrays of solitons for this article beam-to-beam distance
of 80\,\umum has been set, which corresponds to about 3 to 4 beam
diameters.  
This is a compromise between close packing and avoidance of strong 
attracting forces . 
Due to the large distance, the tails of the electric field in the
overlapping region of neighbouring beams have dropped down enough to 
provide for seperate solitary waveguides for every single beam. 
Along with the anisotropy of the refractive index modulation the
waveguiding channels of gradient index type do only give rise to
relevant cross-coupling for much smaller distances of around 2
beam diameters where substantial interaction as well is present.
}%

\subsection{Frequency analysis}

\begin{figure}[htbp]
  \includegraphics[width=\linewidth]{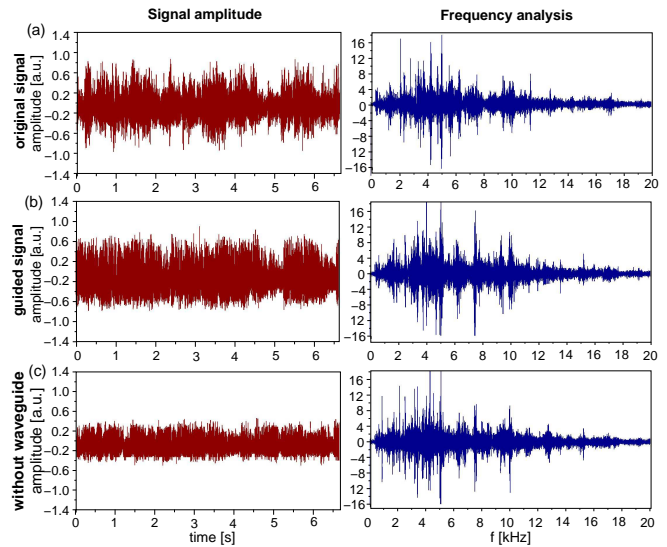}
  \caption{\label{fig:frequency}Frequency analysis: 
    The original signal used to modulate the guided laser beam is
    depicted in (a).
    In the middle row (b) the recorded signal guided in the solitary
    waveguide is shown, which represents the same part of the piece
    of music as in (a). 
    Both shows basically also the same Fourier components. 
    In the lower row (c) the case after waveguide removal is drawn.
    The frequency analysis on the right displays in principle the same
    frequency range for all three cases.
  }
\end{figure}

The quality of information transport is also governed by the correct
transmission of the Fourier components of the signal. 
To investigate the influence of the guiding on the signal quality we
plotted each the amplitude and frequency analysis of the analyzed
signals in figure~\ref{fig:frequency}.
For comparison in the amplitude and frequency for a short part of the
original signal used to modulate the guided laser beam is shown in the
upper row (a). 

The recorded signal guided in the waveguide is depicted in the middle
row (b). 
It represents the same portion of the piece of music as in
(a) and the frequency plot does not show any significant difference. 
Compared with the the third row (c) where the recorded signal after
waveguide removal is shown the amplitude of the guided signal (b) is
about twice as large. 
Although the two parts represent two different positions in the
transmitted piece of music the frequency range and the Fourier
components are basically the same (right column in
figure~\ref{fig:frequency}~(b) and (c)).   
Because the signal-to-noise ratio is different due to the lower signal
amplitude with removed waveduide, the components of the noise are
stronger present. 
The larger dynamic range of the original signal (a) is visible in the
amlitude plot. 
It is limited for the transmitted signal in the experiment due to
modulation and demodulation losses and the higher noise level.

As a conclusion we can state that guiding in photorefractive
soliton waveguides does not show significant influence on the 
characteristics of dynamically amplitude modulated signals. 
The principle suitability of single photorefractive solitons and
complex arrays of these solitons for parallel optical data
communication is therefore demonstrated.

\section{Conclusion}

In this article we have shown successful transmission of dynamically
modulated signals through solitary waveguides in photorefractive
media. 
Analog modulation with an audio signal has been applied using a
magneto-optical amplitude modulator setup and standard personal
computer sound equipment has been used for analyzing the output. 
The short transmission distance of about 22\,mm waveguide length
demonstrates the principle adequacy of spatial photorefractive
solitons to act as waveguides for dynamically amplitude modulated
signals in different wavelength regions.
The signal-to-noise ratio is suffiently large to investigate the
features of the system. 
We did not observe significant principle influences on the
transmission properties of modulated signals in our system.

\mymark{%
Parallel transmission of modulated signals in
neighbouring solitary waveguides has been demonstrated to be a novel
tool for all-optical systems.  
}%
Steered interaction and fusion~\cite{PDSK2002} in combination with
waveguiding in red and infrared wavelength region~\cite{TSSDx2003} are
the other necessary building blocks needed for integration of
all-optical switching devices in telecommunication applications using
optical spatial solitons in photorefractive media.
As a result all-optical data communication more and more gets a
possible application of devices based on photorefractive solitons.

\section*{Acknowledgment}

The authors like to thank Prof. Dr. Horst~D\"otsch from Fachbereich
Physik, Universit\"at Osnabr\"uck, Germany, for the kind donation of
magnetooptic crystal samples used to build the amplitude modulator. 
D.T. acknowledges support from Konrad-Adenauer-Stiftung e.V. 
Partial support from the Graduate School ''Nichtlineare
kontinuierliche Systeme'' by DFG at Westf\"alische
Wilhelms-Universit\"at, M\"unster has been granted.

\bibliography{lit_datensolis}
\bibliographystyle{unsrt}

\vfill

\end{document}